\documentclass[copyright,creativecommons]{eptcs}

\usepackage[colorinlistoftodos,textsize=footnotesize,textwidth=3.3cm]{todonotes}

\providecommand{\event}{9th International Workshop on\\Developments in Computational Models DCM 2013} 
\usepackage{colortbl} 
\usepackage{xspace}
\usepackage{mathpartir}
\usepackage{amsthm}

\newcommand{\til}[1]{\widetilde{#1}}
\newcommand{\gridprocess}[0]{\textsf{Grid\,}^{ \omega,\mu}_{ \delta,\eta}}
\newcommand{\userprocess}[0]{\textsf{User}}
\newcommand{\monitorprocess}[0]{\textsf{UsrMonitor}} 
\newcommand{\approcess}[0]{\textsf{AP}}
\newcommand{\adprocess}[0]{\textsf{AD}}
\newcommand{\usrhdlprocess}[0]{\text{\textsf{AP-UsrHandler}}}
\newcommand{\prxhdlprocess}[0]{\text{\textsf{AP-ProxyHandler}}}
\newcommand{\logprocess}[0]{\text{\textsf{AP-Log}}}
\newcommand{\uprxprocess}[0]{\text{\textsf{AP-UserProxy}}}
\newcommand{\rsrprxprocess}[0]{\text{\textsf{AD-ResourceProxy}}}
\newcommand{\resourceprocess}[0]{\text{\textsf{AD-Resource}}}
\newcommand{\searchprocess}[0]{\text{\textsf{AP-Search}}}
\newcommand{\accprocess}[0]{\text{\textsf{AP-Acc}}}
\newcommand{\receptorprocess}[0]{\text{\textsf{AD-RecReq}}}
\newcommand{\assignprocess}[0]{\text{\textsf{AD-AsgRes}}}
\newcommand{\lrmprocess}[0]{\text{\textsf{AD-LRM}}}

\newcommand{\zero}{\ensuremath{\mathbf{0}}}
\newcommand{\taskprocess}[0]{\lceil \texttt{T} \rceil^{t,e}}

\usepackage{stmaryrd}
\newcommand{\encp}[2]{\llbracket #1 \rrbracket_{#2}}

\newcommand{\name}[1]{\mbox{{{(#1)}}}}
\newcommand{\pired}{\ensuremath{\longrightarrow}}
\newcommand{\wred}{\Longrightarrow}
\newtheorem{defin}{Definition}[section]

\newcommand{\myeqdef}{\ensuremath{\stackrel{\text{def}}{=}}}

\def\midd{\; \; \mbox{\Large{$\mid$}}\;\;}
\newcommand{\para}{\mathord{\;\boldsymbol{|}\;}}

\usepackage{float}
\floatstyle{ruled}
\restylefloat{figure}

\newcommand{\hopi}{\ensuremath{\text{HO}\pi}\xspace}

\usepackage{enumerate} 
\usepackage{breakurl}            
\usepackage{amsmath,amsfonts,amssymb,amsthm,epsfig,epstopdf,url,array}

\title{Towards Formal Interaction-Based Models of \\ Grid Computing Infrastructures}
\author{Carlos Alberto Ram\'irez Restrepo
\institute{EISC -- Universidad del Valle, Cali, Colombia}
\and
Jorge A. P\'{e}rez
\institute{CITI/DI -- FCT - Universidade Nova de Lisboa, Portugal}
\and
Jes\'us Aranda
\institute{EISC -- Universidad del Valle, Cali, Colombia}
\and
Juan Francisco D\'iaz-Frias
\institute{EISC -- Universidad del Valle, Cali, Colombia}
}

\def\titlerunning{Towards Formal Interaction-Based Models of Grid Computing Infrastructures}
\def\authorrunning{C. Ram\'{i}rez \& J. A. P\'{e}rez \& J. Aranda \& J. F. D\'{i}az}

\begin{document}
\maketitle

\begin{abstract}
Grid computing (GC) systems are
 large-scale virtual machines, built upon a massive pool of resources (processing time, storage, software)
that often span multiple distributed \emph{domains}.
Concurrent users interact with the grid by adding new tasks; 
the grid is expected to assign resources to tasks in a fair, trustworthy way. 
 These distinctive features of GC systems make their 
 specification and verification a challenging issue.
Although prior works have proposed formal approaches to the 
specification  of GC systems, a precise account of the 
\emph{interaction model} which underlies  resource sharing 
has not been yet proposed.
In this paper, we describe ongoing work aimed at filling in this gap.
Our  approach relies on \emph{(higher-order) process calculi}: 
these core  languages for concurrency offer a  compositional framework in which  GC systems 
can be precisely described and potentially reasoned about. 
\end{abstract}

\section{Introduction}
\paragraph{Context.}
Grid computing (GC in the following) 
systems 
comprise a large pool of computational resources, 
which are made available by 
 multiple institutions (\emph{administrative domains}) 
 to users wishing to execute tasks
that would be hard (or even impossible) to perform in a single administrative domain.
This is in sharp contrast with usual distributed systems, in which each resource is owned and controlled by a single institution. 
That is, while in distributed systems there is a clear correspondence between system users and valid resource users, 
in GC systems an analogous correspondence is less explicit, as resources may belong to multiple administrative domains. Moreover, 
a grid user may not correspond to an actual user in the administrative domains. 
Yet another point of contrast concerns 
transparency and security requirements: 
while in conventional distributed systems users typically know a priori the resources that they need for executing their tasks, 
grid users may execute tasks without being aware of the internal structure of the system. 
GC systems differ also from emerging cloud computing platforms, 
which offer economies of scale for exploiting virtually unlimited resources, based on the Software as a Service (SaaS) paradigm. 
In fact, differently from clouds, GC systems aim at  executing computationally intensive tasks, subject to 
constraints on 
resource availability/access.
Other notable differences between clouds and grids concern failure management, resource ownership, and infrastructure transparency~\cite{Foster360}.



%

\paragraph{This Work.}
Here we are concerned with 
principled approaches to 
 the correct design and construction of GC systems.
As discussed above, 
a critical aspect is that of
appropriately assigning resources to a potentially huge number of user tasks running concurrently.
Given the scale, complexity, and peculiarities of GC systems, this is  
a challenging issue from several perspectives. 
In this paper, 
we describe 
ongoing work aimed at tackling this issue from the perspective of 
formal models of computation based on \emph{communication}. 
More precisely, we explore a \emph{process calculi} approach: 
based on a small set of operators ---typically, atomic interaction, sequencing, parallel composition, and scoping--- process calculi such as  CCS~\cite{Milner89} and the $\pi$-calculus~\cite{MilnerPW92a} have been developed within the concurrency theory community as basic models for 
communicating systems.
As process calculi are \emph{compositional}, they have proved useful for developing reasoning techniques over  specifications (e.g. behavioral equivalences and type systems) and for investigating new programming abstractions based on communication.
These features make process calculi an attractive basis for 
the formal specification and verification of GC systems.


%

In particular, in this paper we rely 
on \emph{higher-order  process calculi}, i.e.,
calculi 
in which processes (more generally, values containing processes) can be communicated.
This is in contrast to 
\emph{first-order} calculi such as the $\pi$-calculus, 
in which only basic values and/or 
communication channels
can be exchanged.
Higher-order process calculi can be seen as concurrent variants of the $\lambda$-calculus.
In fact, the reduction rule for  communication in these formalisms is reminiscent of well-known $\beta$-reduction in functional calculi.
In the grid setting, 
higher-order 
(or process passing) concurrency naturally models the fact that user tasks ---typically, arbitrarily complex descriptions of computational behaviors--- need to be 
exchanged among different grid components in order to achieve their execution.
In particular, we rely on the higher-order $\pi$-calculus (\hopi)~\cite{sangiorgi:highorder}, a 
core language  which 
enhances the name passing abilities of the $\pi$-calculus with process passing. 
\hopi specifications can represent forms of code mobility, 
therefore allowing for flexibility in descriptions of concurrent communications. 
Moreover, useful proof techniques based on behavioral equivalences are well-understood for (variants of) \hopi (see, for instance,~\cite{DBLP:journals/iandc/Sangiorgi96}).

The main contribution of this work is a formal model of GC infrastructures,
with a focus on the resource assignment facility that is central to them.
Our model distills the main features of GC systems, as informally discussed in the literature (see, e.g., \cite{foster_kesselman_tuecke:grid_ana})
and as identified in our own exchanges with GC experts.
The model is divided into \emph{static} and \emph{dynamic} components. 
The static component, defined in first-order logic,
formalizes the essential pre-conditions and invariants
that should hold for the different grid subsystems.
Using \hopi processes, the dynamic component 
captures the (concurrent) 
execution sequences associated to potentially many users interacting simultaneously with the grid.
These components are intended to be complementary: building upon the relations defined by the static component, 
the dynamic component accounts for the main agents present in real GC infrastructures, 
such as users, tasks, administative domains, virtual organizations, and  resources. 
Our model also considers user and resource proxies, which facilitate user interaction with the GC system and resource management (see~Sect.~\ref{sec:grid}).

While simple, our formal model 
already provides a good basis for obtaining more detailed descriptions of GC systems
and for reasoning about their correctness properties. 
Examples of such  properties 
are authentication and authorization guarantees: they are 
intended to ensure that users only access and use the administrative domains and resources for which they can prove their identity/permissions.
An extension of our current model with suitable cryptographic elements
(using, e.g., the higher-order language in~\cite{DBLP:conf/esop/KoutavasH11}) would be a  step in this direction.
Another relevant property 
concerns the balanced assignment of administrative domains and their use of resources. 
To this end, process specifications of different scheduling and assignment  policies may be necessary ---this issue is largely orthogonal to the model given here.
Other properties of interest for GC modelers involve task termination and resource delivery aspects. 
By  adapting known results on termination and reachability properties 
for calculi such as CCS~\cite{DBLP:journals/mscs/BusiGZ09} and for variants of \hopi~\cite{DBLP:journals/iandc/LanesePSS11,DBLP:conf/ictac/GiustoPZ09}, our model could offer an alternative for investigating such properties.


\paragraph{Related Work.}
We believe that our work improves on previous attempts for grid formalization.
The $\pi$-calculus has been used in~\cite{weng_lu_deng:formservicepubgrid,zhou_zeng:gridservcomp_mech,zhanjun_yongzhong_shaozhong:picalculusgridworkflow} for  
analyzing the specific aspects of grid services composition and workflow. These approaches only model the GC components related to grid services such as resources 
and tasks; other aspects of the GC dynamics are not considered.
In contrast, our model adopts a more comprehensive view of GC systems, including, e.g.,
 key interaction patterns related to user intervention, and the r\^{o}le of user and resource proxies in resource assignment and task execution.
 In~\cite{nemeth_sunderam:charecgrid}, 
 Abstract State Machines (ASM) are used to
give a declarative 
characterization of GC systems; this characterization  
formally describes some of the main attributes that a GC system should support. 
The GC elements are modeled as universes (sets); their behavior is represented using rules over universes.
The only grid agents considered in~\cite{nemeth_sunderam:charecgrid} are tasks (there called \emph{processes});  there are also user and resource mapping agents. Each agent executes the rules over the defined universes.
In contrast to our work,
in the model of~\cite{nemeth_sunderam:charecgrid}
concurrent interactions among GC components are not explicitly represented; 
also, 
such a model
does not consider the key concept of virtual organizations and the r\^{o}le of user and resource proxies.
Finally, in~\cite{bratosin_aalst_sidorova_nikola:referencemodelgrid,du_jiang_guo:formalmodelgridarchpetri}, 
high-level and colored Petri nets were used to analyze grid architectures and grid workflows. 
A 3-layer grid architecture and the interaction between GC components in these layers is modeled.
However, these approaches do not consider virtual organizations, administrative domains, and security requirements ---all of these being central elements in resource assignment.

\paragraph{Organization.} The rest of this paper is organized as follows. Sect.~\ref{sec:grid} briefly recalls the main features of GC systems. In Sect.~\ref{sec:process} we present the syntax and semantics of the higher-order $\pi$-calculus. 
Sect.~\ref{sec:model} gives a brief description of our GC formalization, and 
Sect.~\ref{sec:example} illustrates it via a small example.
Finally, future work is discussed in Sect.~\ref{sec:future}.
A full description of our formal process model is available in~\cite{our:TR}.

\section{Grid Computing: A Brief Overview}
\label{sec:grid}

Grid computing broadly refers to the coordinated resource sharing and problem solving in dynamic, multi-institutional virtual organizations. 
GC systems often require interoperability features and support for heterogeneous environments.
Other typical requirements are 
decentralized control, 
security, access transparency, scalability, availability, and reconfigurability \cite{foster_kesselman_tuecke:grid_ana}.
Sharing in GC systems not only refers to data and information but also to direct access to all kinds of resources 
which may be required for executing complex tasks
(computing power, storage, software applications, data). 
Each \emph{administrative domain} 
(AD in the following)
establishes  what resources are shared  and
their access and usage policies. A \emph{virtual organization} (VO in the following) is a set of ADs defined by such policies. 
The participants in a VO share resources in a controlled way in order to cooperate in executing a specific task.
VOs vary in their purpose, scope, size, duration, structure, community, and sociology~\cite{foster_kesselman_tuecke:grid_ana}.

In GC systems, users can transparently share or access resources---they do not need to know (or be aware of) what resources they are using,
where such resources are physically located or that they may have previously failed and recovered. 
This transparency is achieved by the so-called \emph{grid middleware}.
This is a software layer 
that (i) implements the \emph{protocols} and \emph{services} that 
 enable the seamless sharing of heterogeneous resources, 
and (ii) provides key functionalities for enabling task execution and VOs establishment~\cite{stanoevska_wozniak_ristol:gridcloudcomputing}.
In this way, the middleware allows users to access resources while satisfying security policies  such as authentication, authorization, delegation, and single sign-on. 
To this end, the grid middleware  includes user and resource \emph{proxies}~\cite{foster_kessselman_tsudik_tuecke:security_arch}. While a user proxy is an entity that is given permission to act on behalf of a user for a fixed period of time, a resource proxy  serves as interface between the middleware and a resource, thus simplifying 
(i) the authentication between user proxy and the resource and 
(ii) the mapping between grid users and the local users which are valid in the resource.

\begin{center}
\begin{figure}[t!]
\includegraphics[scale=0.265]{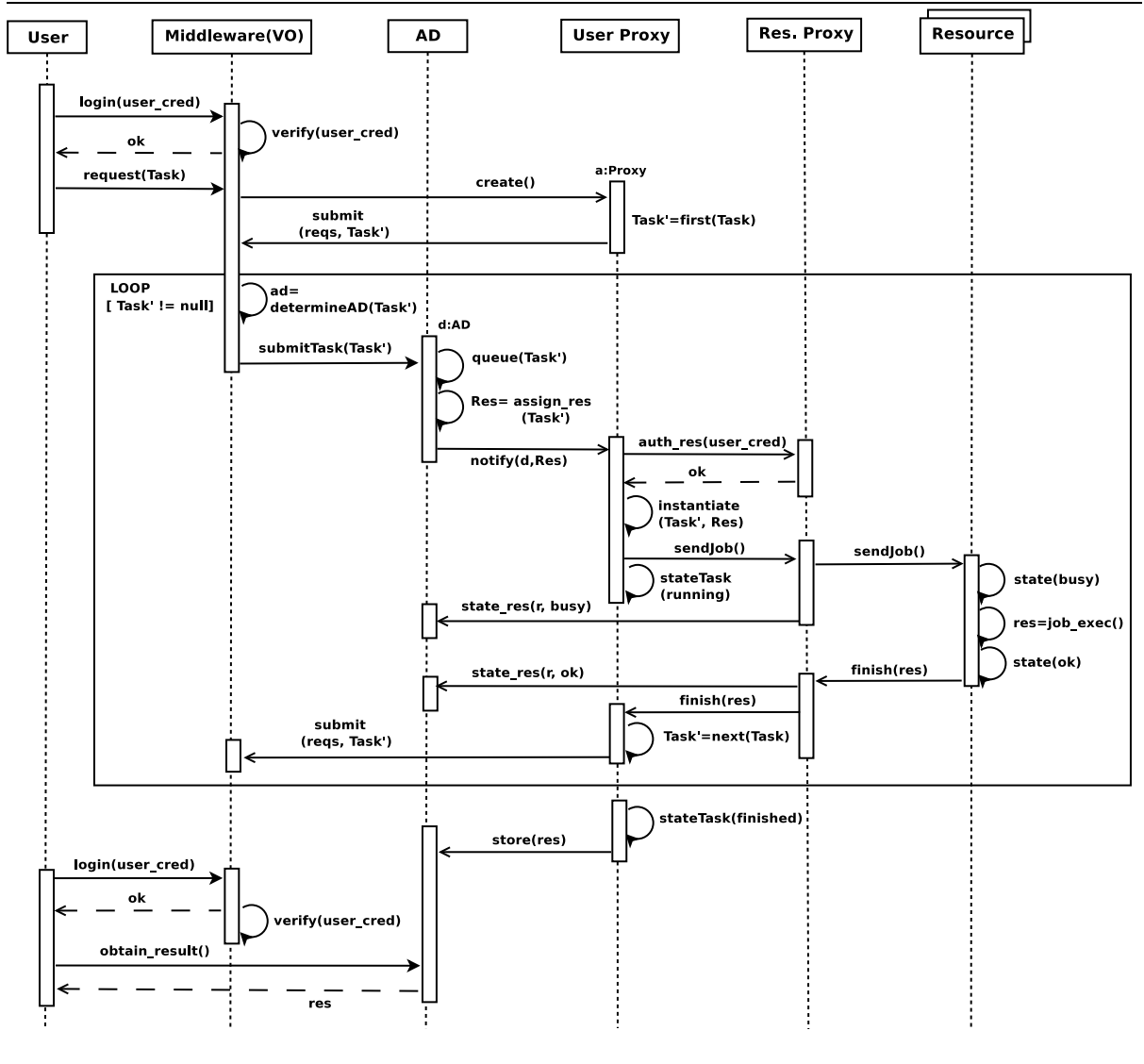}
\caption{The Grid Interaction Protocol as a Sequence Diagram}
\label{fig:protocol}
\end{figure} 
\end{center}

\paragraph*{Grid Resource Assignment Protocol.}
As our interest is in an interaction-based approach to GC systems, below we 
present a protocol which 
describes the interaction sequence among the main grid components (users, ADs, VOs, resources, proxies).
The protocol is described as a sequence diagram in Fig.~\ref{fig:protocol};
it formalizes requirements and mechanisms which have been 
described in the literature only informally. The formal model given in Sect.~\ref{sec:model}
is then intended to give a precise account of this protocol.

\begin{enumerate}[1.]
\item A user sends its credentials to a grid node in order to authenticate. In the figure, 
this step is represented by the message $login(user\_cred)$ from User to VO.

\item If the authentication is successful then the user is granted to access the grid. 
Otherwise, the user must revise its credentials. 
For simplicity, the figure shows only the case in which authentication is successful; this step is represented by message $ok$ from VO to User.

\item The authenticated user sends a proxy creation request, and a task with its requirements to the grid node.
The task may be a complex object; in particular, it may be structured in terms of subtasks which follow some process logic. 
In the figure, these steps are represented by messages $userproxy\_creation()$ (from User to VO), $create()$ (from VO to User Proxy), 
and $request(Task)$ (from User to VO).

\item The user proxy sends to the grid node the requirements of each subtask. 
In the figure, this step is represented by the message $submit(reqs, Task')$ from User Proxy to VO.

\item The  node selects an AD in the VO with available resources to satisfy the subtask requirements. 
This subtask is assigned and sent to the selected AD. In Fig.~\ref{fig:protocol}, this is represented by messages 
$determineAd(Task')$ (inside VO), $submitTask(Task')$ (from VO to AD), and $queue(Task')$ (inside~AD).

\item The AD assigns appropriate resources for this subtask according to some scheduling strategy. 
In the figure, this step is represented by the message $assignRes(Task')$ inside AD.

\item The user proxy authenticates into the resource proxies of assigned resources. 
If authentication is successful then the subtask is executed.
Otherwise, the subtask is sent back to the grid node. 
In the figure, these steps are represented by messages $auth\_res(user\_cred)$ (from User Proxy to Resource Proxy), 
$ok$ (from Resource Proxy to User Proxy), $sendJob()$ (from Resource Proxy to Resource), and $job\_exec()$ (inside Resource).

\item When the subtask has finished (message $finish(res)$, from Resource to its Resource Proxy), it is detected if there are more subtasks (condition $Task'\neq null$ in the loop). If yes then the result of the previous subtask is transmitted to the next subtask and the 
previous subprotocol is executed again (message $submit(reqs,Task')$). 
Otherwise, if the just executed subtask is the last one, then the result is stored and the protocol finishes (message $store(res)$ from User Proxy to AD).

\end{enumerate}

\paragraph*{A Representative GC Scenario.}

\begin{figure}[t!]
\begin{center}
\includegraphics[scale=0.36]{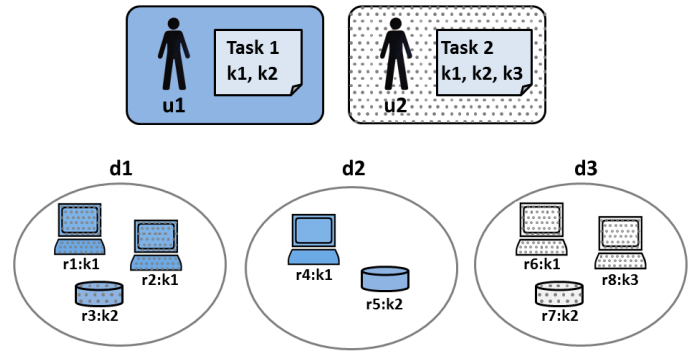}
\end{center}
\caption{A GC Scenario: Two users ($u_1$, $u_2$), two VOs ($v_1$ -- blue, $v_2$ -- dotted), three ADs ($d_1$, $d_2$, $d_3$)}
\label{fig:samplegrid}
\end{figure}

We now describe a small, representative example of a GC system. 
Depicted in Fig.~\ref{fig:samplegrid}, our scenario draws 
inspiration from the one given in \cite{foster_kesselman_tuecke:grid_ana}.
It contains three ADs (denoted $d_1$, $d_2$, and $d_3$) and two VOs (denoted $v_1$ and $v_2$);
in the figure, they are depicted as ovals and rectangles, respectively. 
VO~$v_1$ (blue background) groups participants in an aerospace design consortium and $v_2$ (dotted background) links participants for sharing spare computing cycles. AD $d_1$ is member of both $v_1$ and $v_2$. Also, ADs $d_1$ and $d_2$ participate in $v_1$ and AD $d_3$ participates in $v_2$.
We also consider  users  $u_1$ and $u_2$:
while $u_1$ belongs to $v_1$, user $u_2$ belongs to $u_2$. 
Both $u_1$ and $u_2$ have a task to execute in the grid, denoted Task1 and Task2 in the figure, respectively. 
To perform, Task1 requires one resource of type (or descriptor) $k_1$ and one resource of type $k_2$. 
Similarly, Task2 requires three resources, distinguished by types $k_1$, $k_2$, and $k_3$.
Resources are located in appropriate ADs:
AD $d_1$ owns three resources: $r_1$ (type $k_1$), $r_2$ (type $k_1$), and $r_3$ (type $k_2$); 
AD $d_2$ owns two resources: $r_4$ (type $k_1$) and $r_5$ (type $k_2$); and
AD $d_3$ owns three resources: $r_6$ (type $k_1$), $r_7$ (type $k_2$), and $r_8$ (type $k_8$).
While resources of $d_1$ are shared by $v_1$ and $v_2$, resources of $d_2$ are available only to $v_1$,
and resources of $d_3$ are available only to $v_2$.
%

\section{The Process Model: Syntax and Semantics}
\label{sec:process}

This section briefly presents the syntax and semantics of 
the higher-order $\pi$ calculus,  \hopi. 
Our presentation closely follows~\cite{sangiorgi:highorder}.
In \hopi, both names (communication channels) and processes may be passed around by synchronization on names;
communication can be thus loosely assimilated to $\beta$-reduction in the $\lambda$-calculus.
We assume a set of names/channels ranged over $x,y,z,\ldots$ and a set of process variables ranged over $X,Y,Z,\ldots$. 
We write $\til{o}$ to denote a finite tuple of elements $o_1, \ldots, o_k$.

\begin{defin}
The language of \hopi \emph{processes} is 
given by the following syntax:
\begin{eqnarray*}
 \alpha  &::= &   x(\til{U})   \midd   \overline{x}\langle \til{K} \rangle \\
P      &::= &  \sum\limits_{i\in I}{\alpha}_i.P_i   \midd    P_1 \para P_2  \midd    (\nu\ x)P  \midd    \mathtt{if}\ [x=y]\  \mathtt{then}\ P_1\ \mathtt{else}\ P_2   \midd    \mathsf{D}\langle \til{K} \rangle   \midd    X\langle \til{K} \rangle 
\end{eqnarray*}
\end{defin}
We have two \emph{prefixes}, ranged over $\alpha, \alpha', \ldots$.
An input prefix $x(\til{U})$ (resp. output prefix $\overline{x}\langle \til{K} \rangle$)
denotes an atomic input action (resp. output action) on a name $x$. 
Above, $\til{K}$ and $\til{U}$ 
denote tuples of names and processes, and of names and variables, respectively. 
Process $\sum\limits_{i\in I}{\alpha}_i.P_i$ represents the \emph{non-deterministic choice} among prefixed processes ${\alpha}_i.P_i$. 
The operational semantics ensures that only one of them will be executed, discarding the rest. 
When $I = \emptyset$ we write $\zero$; when $I = |2|$ we write $\alpha_1.P_1 + \alpha_2.P_2$.
Also, we simply write $\alpha$ to refer to process $\alpha.\zero$.
Process $P_1 \para  P_2$ stands for the \emph{parallel composition} of processes $P_1$ and $P_2$. 
We write $\prod\limits_{j\in J} P_j$ as a shorthand notation for process $P_1 \para \ldots \para P_{|J|}$.
Process $(\nu x)P$ declares the name $x$ private to process $P$. That is, the scope of $x$ is $P$; this scope may be enlarged 
by communication to other processes (\emph{scope extrusion}).
The conditional $\texttt{if}\ [x=y]\  \texttt{then}\ P_1\ \texttt{else}\ P_2$ is 
based on equality of names $x$ and $y$: if $x= y$ then the process continues as $P_1$; otherwise it continues as $P_2$.
By taking inputs and restriction as binders, notions of free and bound names/variables arise as expected.
We identify processes up to consistent renaming of bound names/variables, writing $\equiv_\alpha$ for this congruence.

One way of specifying infinite process behavior is via \emph{parametric definitions}.
Notation $\textsf{D}\langle \tilde{K} \rangle$ denotes the application of a constant identifier $\textsf{D}$ 
with parameters $\til{K}$.
We assume each   $\textsf{D}$ has a unique definition $\textsf{D}(\tilde{U})\ \stackrel{def}{=}\ P$, where 
$\til{U}$ is composed of all free names or variables in $P$, i.e. names or variable which occur out the scope of any binding.
Then, $X\langle \tilde{K} \rangle$ denotes the application of parameters $\til{K}$ to process  variable $X$. 

We endow our process language with a 
\emph{reduction semantics}. 
Intuitively, a reduction $P \longrightarrow Q$ denotes a single evolution step from process $P$ to $Q$,  without interaction from its surrounding environment. 

\begin{defin}
\emph{Reduction},  $P \pired Q$, is the binary relation on  processes 
defined by the rules in Fig.~\ref{fig:red_sem}.
\end{defin}

\begin{figure}
\begin{center}
\vspace{4mm}
\begin{tabular}{c}
\inferrule[\name{$\mathsf{COM}$}]{} {(\ldots  + x(\til{U}).P)  \para   (\ldots + x\langle \til{K} \rangle.Q) \pired P\{ \til{K}/\til{U} \} \para Q} \vspace{4mm} \\ 
\inferrule[\name{$\mathsf{PAR}$}]{P \pired P'}{P \para Q \pired P'  \para Q} \qquad 
\inferrule[\name{$\mathsf{RES}$}]{P \pired Q}{(\nu x)P  \pired (\nu x)Q}  \qquad    
\inferrule[\name{$\mathsf{STR}$}]{P \equiv P' \quad P' \pired Q' \quad Q' \equiv Q}{P  \pired Q} 
\end{tabular}
\end{center}
\caption{Reduction semantics for $\hopi$}
\label{fig:red_sem}
\end{figure}

As usual, we write $\wred$ to denote the reflexive, transitive closure of $\pired$.
The rules in Fig.~\ref{fig:red_sem} formalize process communication and reduction under parallel composition and restriction. 
In  rule \name{$\mathsf{COM}$}, notation
$P\{ \til{K}/\til{U} \}$ stands for process $P$ in which 
all free occurrences of names/variables in $\til{U}$ 
have been substituted by names/processes in $\til{K}$. We assume arity in communications is consistent, i.e.,
the length of $\til{U}$ must be equal to the length of $\til{K}$, with one-to-one correspondence among elements of both tuples.
By means of rule \name{$\mathsf{STR}$}, reduction is closed under a \emph{structural congruence} relation, written $\equiv$, which is used to promote process interactions. It is defined as follows:


\begin{defin}
Structural congruence, written $P \equiv Q$, is the  smallest process congruence 
such that
  $$
\begin{array}{c}
  P \para \zero  \equiv P \qquad
  P \equiv_{\alpha} Q \Rightarrow P \equiv Q  \qquad
  P \para Q \equiv Q \para P  \qquad
  P \para (Q \para R) \equiv (P \para Q) \para R  \qquad
  (\nu x)\zero \equiv \zero \\
  x\not\in fn(P) \Rightarrow P \para (\nu x)Q \equiv (\nu x)(P \para Q)   \qquad
     (\nu x)(\nu y)P \equiv (\nu y)(\nu x)P \\
 \mathtt{if}\ [x=y]\ \mathtt{then}\ P_1\ \mathtt{else}\ P_2\ \equiv ~P_1 ~(\text{if $x=y$}) \qquad 
   \mathtt{if}\ [x=y]\ \mathtt{then}\ P_1\ \mathtt{else}\ P_2\ \equiv ~P_2 ~(\text{if $x\neq y$})  \\
  \textsf{D}(\til{U}) \myeqdef P ~~\Rightarrow~~ \textsf{D}\langle \til{K} \rangle\ \equiv \ P\{ \til{K}/\til{U} \} \qquad \sum\limits_{i\in I} {\alpha}_i.P_i\ \equiv\ \sum\limits_{j\in J} {\alpha}_j.P_j  ~(\text{if $J$ is a permutation of $I$}) 
  \end{array}$$
\end{defin}

\section{A Formal Model of Grid Interaction}
\label{sec:model}
We now give an overview of our formal model of GC systems; a full description can be found in~\cite{our:TR}.
The model is intended as a formal counterpart of the informal interaction protocol given in Sect.~\ref{sec:grid}.
As already discussed, the model is divided into static and dynamic components. 
While the former is given in terms of invariants (first-order logic formulas),
the latter is specified using \hopi processes.
The two components play complementary r\^{o}les in our model.
On the one hand, the invariants and conditions in the static part are used to:
\begin{enumerate}[$-$]
\item Define the actors in the system (e.g. users, administrative domains, resources, tasks) and useful relationships between them; 
\item Describe the initial configuration of the system;
\item Define well-formedness conditions for the processes of the dynamic part.
\end{enumerate}
On the other hand, the dynamic part focuses on representing: 
\begin{enumerate}[$-$]
\item How the grid assigns resources to each task;
\item The start of task execution; 
\item The state of tasks and their assigned resources.
\end{enumerate}
It is worth highlighting that processes of the dynamic component cannot add new tasks, resources or users. Although these capabilities are present in some real grid systems, in the current development we focus on systems in which those elements cannot be added at runtime.

\subsubsection*{Static Component: Base Sets and Invariants}


In order to 
formalize the key components of GC systems, we first relate such components to base reference sets. 
Then, we state associated invariant properties by defining static predicates over elements of such sets.
Table~\ref{table:base_sets} summarizes our notation for these base sets.
The intuitive meaning of most of them should be clear from the description given in Sect.~\ref{sec:grid}.
We consider that each VO is associated to a group of access points (nodes) which are contained in the base set $N$.
Observe that we distinguish between  \emph{task definitions} (which belong to base set $T$) and  \emph{task instances}, which are submitted by users (and belong to base set $S$). 
We assume that task definitions are built using the next grammar:
\begin{equation*}
\mathtt{T} ::= 
\mathtt{J} \langle k_1,\ldots,k_m \rangle  
 \midd   \mathtt{T}.\mathtt{T}  \midd      \mathtt{T}\parallel \mathtt{T}  \midd   \mathtt{T}  \oplus  \mathtt{T} \midd    \mathtt{end}  
\end{equation*}

\noindent Above,
 $\mathtt{J} \langle k_1,\ldots,k_m \rangle$ 
denotes a \emph{basic task} $\texttt{J}$ with resources of type $k_1,\ldots,k_m$, respectively ($m \geq 1$). 
Building upon basic tasks, 
more complex ones can be defined, using sequential and parallel composition (denoted $\mathtt{T}.\mathtt{T}$ and $\mathtt{T} \parallel \mathtt{T}$, respectively)
and non-deterministic choice ($\mathtt{T}  \oplus  \mathtt{T}$). We also assume a termination task, denoted $\texttt{end}$.
As discussed above, we assume that each user is associated to a single task. 
This is not a limitation, for tasks may involve several subtasks in parallel and sequential composition.

As for the invariants, 
based on informal descriptions in the literature~\cite{foster_kesselman_tuecke:grid_ana}, 
we have identified the elements that we consider essential to GC systems. 
Using first-order logic, 
we formalize such elements in terms of  predicates over the elements of the reference sets. 
Some of such invariants are the following:

\begin{figure}[t]
\begin{center}
\small
\begin{tabular}{|l|l||l|l|}\hline
\textbf{GC Component} & \textbf{Base set} & \textbf{GC Component} & \textbf{Base set} \\\hline
Users                 & $u\in U$  & VOs                   & $v\in V$  \\\hline
ADs                   & $d\in D$  & Tasks                 & $\mathtt{T}\in T$  \\\hline
Resources             & $r\in R$  & User Tasks            & $\mathtt{S}\in S$    \\\hline
Nodes                 & $n\in N$  & User Proxies          & $a\in A$  \\\hline
Resource Proxies      & $x\in X$  & Resource Descriptors           & $k\in K$  \\\hline
Logs                  & $l\in L$  & &\\\hline
\end{tabular}
\caption{Static model: Base sets for GC components}\label{table:base_sets}
\end{center}
\end{figure}

\begin{enumerate}[$-$]

\item \emph{Each user is member of exactly one VO.} Using
predicate $ member(u,v)$, which holds if user $u\in U$ is member of VO $v\in V$, we  may state this invariant as:
\begin{center}
$\forall_{u\in U}, \exists_{v\in V}.\ member(u,v)\ \wedge\ \forall_{u\in U,\ v,v'\in V}.\ (member(u,v)\ \wedge\ member(u,v')\ \rightarrow\ v=v')$
\end{center}

\item \emph{Each user is associated to exactly one task to be executed in the GC system.} Using predicate $task(u,\texttt{S})$,
which holds if user $u\in U$ is the owner of task $\texttt{S}\in S$, we may state this invariant as:

\begin{center}
$\forall_{u\in U}, \exists_{\texttt{S}\in S}.\ task(u,\texttt{S})\ \wedge\ \forall_{u\in U,\ \mathtt{S},\mathtt{S'}\in S}.\ (task(u,\mathtt{S})\ \wedge\ task(u, \mathtt{S'})\ \rightarrow\ \mathtt{S}=\mathtt{S'})\ \wedge\ $\\
$\forall_{u,u'\in U,\ \mathtt{S}\in S}.\ (task(u,\mathtt{S})\ \wedge\ task(u',\mathtt{S})\ \rightarrow\ u=u')$.
\end{center}

\item \emph{Each resource belongs to exactly one AD.} Using predicate $belongsTo(r,d)$, which holds if resource $r\in R$ belongs to AD $d\in D$, we may express this invariant as:
\begin{center}
$\forall_{r\in R}, \exists_{d\in D}.\ belongsTo(r,d)\ \wedge\ \forall_{r\in R,\ d,d' \in D}.\ (belongsTo(r,d)\ \wedge\ belongsTo(r,d')\ \rightarrow\ d=d')$
\end{center}

\item \emph{Every AD can participate in one or more VOs.} Using predicate $participate(d,v)$, which holds 
if AD $d\in D$ participates into VO $v\in V$, we may state this invariant as:
$\forall_{d\in D}, \exists_{v\in V}.\ participate(d,v)$.

\end{enumerate}

Additional invariants concern
access points (nodes), resource descriptors, task states, resource states, and task logs;
they are given in terms of appropriate base sets, 
and are omitted here for the sake of space. 

\subsubsection*{Dynamic Component: Model in the $\hopi$ calculus}

\begin{figure}[t]
\begin{center}
\small
\begin{tabular}{|c|l|l|}\hline
\textbf{GC Component} & $\mathbf{\hopi}$ \textbf{Process} & \textbf{Intuitive Description} \\\hline
Grid system          & $\gridprocess$     & Represents the whole GC system    \\\hline
User ($u\in U$)      & $\encp{u,\mathtt{S}}{}^{\til{c},y}=\userprocess(\til{c}, \encp{\mathtt{S}}{}^{t,e}, y)$ & Models the behavior of $u$ to authenticate \\
                      &                                   & and submit its task $\mathtt{S}$ \\\cline{3-3}
                      & $\ \ \ \ \ \ \hookrightarrow \monitorprocess(\til{c},g,a,y,\textsf{P})$  & Monitors tasks submitted by $u$ \\\hline 
Node ($n\in N$)      & $\encp{n}{}^{v,y,\til{d}}=\approcess(y, \til{d})$  & Models the interaction of $n$ with users \\
                      &                                 & to authenticate\\
                      & $\ \ \ \ \ \ \hookrightarrow\usrhdlprocess(ch_1, ch_2, ce)$  & Models the user proxy creation and task\\
                      &                                  & submission               \\
                      & $\ \ \ \ \ \ \hookrightarrow\prxhdlprocess(ce, \til{d})$ & Represents the interaction with the user task \\\hline
VO ($v\in V$)        & Composition of instances of $\approcess(y, \til{d})$ & A collection of nodes \\\hline
AD ($d\in D$)        & $\encp{d}{}=\adprocess(d)$      &  Models the AD with its resources, proxy \\
                      &                    & resources and management elements \\\cline{3-3}
                      & $\ \ \ \ \ \ \hookrightarrow\receptorprocess(b,d)$ & Receives the tasks assigned to the AD and \\
                      &                 & puts them in the queue \\\cline{3-3}
                      & $\ \ \ \ \ \ \hookrightarrow\assignprocess(b,d,ch)$ & Dequeues tasks and assigns \\
                      &                 & appropriate resources to them \\\cline{3-3}
                      & $\ \ \ \ \ \ \hookrightarrow\lrmprocess(\til{s},\til{x},\til{w},ch,d)$ & Supervises the state of resources, and  \\
                      &                 & determines the available resources for a task \\\hline
Resource ($r\in R$)  & $\encp{r}{}^{r,q}=\resourceprocess(r,q)$  & Models a resource's behavior when is \\
                      &                 & used by a task           \\\hline
User Proxy ($a\in A$) & $\encp{a}{}^{ce,p,t,g}=\uprxprocess(ce,p,t,g)$  & Models the task management, the request of \\
                        &               & execution of subtasks and the authentication \\
                        &               & with resource proxies        \\\hline
Res. Proxy ($x\in X$) & $\encp{x}{}^{x, q, r,w}=\rsrprxprocess(x, q, r,w)$  & Acts as a mediator between GC components  \\
                        &                  & and a resource      \\\hline
Log ($l\in L$)       & $\logprocess(g_r,g_w,st,\til{z})$ & Interacts with GC components to register the\\
                     &                                & changes in the task state and result  \\\hline
Task ($\mathtt{T}\in T$)  & $\taskprocess$ definition & Represents the behavior of a task  \\\hline
User Task ($\mathtt{S}\in S$) & $\encp{\mathtt{S}}{}=\taskprocess$ & Models a task instance corresponding to \\
                               &                & a user task \\\hline
Descriptors ($k\in K$)         & Names $\mathtt{k_1},\ldots,\mathtt{k_\kappa}$ & Models the different types of resources   \\\hline
\end{tabular}
\caption{Dynamic model: Correspondence among GC components and processes (full details in~\cite{our:TR})}
\label{table:stat_dyna}
\end{center}
\end{figure}

In addition to 
specifying the main system components and the valid relations among them, 
our model should unambiguously describe how 
the system may evolve as a result of the interaction of its components.
In the light of the  protocol given in Sect.~\ref{sec:grid},
such interactions may follow intricate patterns and must adhere to 
basic correctness and trustworthiness criteria. 
We would like formal mechanisms to ensure that models indeed satisfy such criteria.
As we wish to describe GC systems compositionally, precisely specifying the interacting mechanisms and their relationships, 
first-order logic is not the most appropriate formalism for this task. 
We then appeal to specifications expressed as \hopi 
processes: they offer a basis on which
interaction features in GC systems can be succinctly represented, and potentially verified using reasoning techniques over interacting  processes. 
We thus extend the static description overviewed above so as to define in \hopi the behavior of GC components and their interactions according to the invariants and predicates of the static representation. Fig.~\ref{table:stat_dyna} summarizes the correspondence between the elements in the static description and their respective process representation in the dynamic component of the model. 
In the figure, we use the symbol $\hookrightarrow$ to represent sub-processes which are triggered as part of the execution of a main process.
Complete descriptions of the processes mentioned the figure can be found in~\cite{our:TR}.

%
%

Next we briefly describe process representations for some grid components (users, middleware, ADs) mentioned in 
Fig.~\ref{table:stat_dyna}.
We use $\omega$, $\mu$, $\delta$, and $\eta$ to denote, respectively, the number of VOs, users, ADs, and nodes (access points) in the system.
Also, we rely on standard process representations of queues (and associated operations) 
which can be easily encoded in \hopi via name passing (see, e.g.,~\cite{SaWabook}).
It is worth highlighting that the \hopi representations of the GC components are related to the invariants and other elements of the static component of the model. This means that process interactions 
do not concern arbitrary elements of the base sets; rather, they involve elements which may be subject to invariants.
For example, our process representation for users only can interact with the process representation of a node that corresponds to a VO where such a user is member.   
Interestingly, key elements of the process language (notably, 
the exchange of fresh channels and scope extrusion)
turn out to be useful to enforce the static invariants in the dynamic specification,
and to rule out
undesirable interferences among components (as in, e.g., two users which concurrently access a VO).
This way, sensible correctness/consistency properties are ensured by construction.
Establishing a formal  
correspondence between the static and dynamic components  is part of ongoing work (see Sect.~\ref{sec:future}).


\paragraph{Grid system.}
A grid system is modeled as the  composition of 
processes representation of users, ADs, and access points.
These are denoted $\encp{u, \mathtt{S}}{}^{\til{c},y}$, $\encp{d}{}$, and $\encp{n}{}^{v,y,\til{d}}$, respectively, 
which are used as intermediate notations for processes
$\userprocess(\til{c}, \encp{\mathtt{S}}{}^{t,e}, y)$,
$\adprocess(d)$, and $\approcess(y, \til{d})$, respectively.
This structure promotes interaction:
while user processes interact with access point processes through private channels $y_1,\ldots,y_{\eta}$, 
AD processes communicate with access point processes in private channels $d_1,\ldots,d_{\delta}$. 
This way, our process model of a GC system, parametric on $\omega$, $\mu$, $\delta$, and $\eta$, is the following:
\begin{eqnarray}
\gridprocess      &\myeqdef& (\nu\ y_{n_1}, \ldots, y_{n_\eta}) \big (\prod\limits_{i\in I} \encp{u_i, \mathtt{S_i}}{}^{\til{c_i},\,y_{node(u_i)}}\ \para\ (\nu\ d_1,\ldots, d_{\delta}) (\prod\limits_{h\in H}  \encp{n_h}{}^{vo(n_h), \,y_{n_h}, \,\til{d_h}} \ \para\ \prod\limits_{l\in L} \encp{d_l}{})\big) \nonumber
\end{eqnarray}


where $I=\{1,\ldots,\mu\}, H=\{1,\ldots,\eta\}$, and $L=\{1,\ldots,\delta\}$ are index sets over users, access points, and ADs, resp. 
In process $\userprocess(\til{c}, \encp{\mathtt{S}}{}^{t,e}, y)$ (defined below),
$\encp{\mathtt{S}}{}^{t,e}$ is a process representation of task $\texttt{T}$ (where $\mathtt{S}$ is a instance of $\mathtt{T}$) that depends on names $t$ and $e$: while $t$ is used to send subtasks requirements to the appropriate user proxy, $e$ is used to signal task completion.
Given $i\in I$, we write 
 $node(u_i)$ to denote the index of the access point for user $u_i$.
Name $d$ in $\adprocess(d)$ is used 
for interaction between the AD and access point processes. 
In $\approcess(y, \til{d})$, name $y$ is used to interact with user processes, while  $\til{d}$ stands for a tuple with the access channel of the ADs in the VO associated to the access point. We write $vo(n_h)$ to denote the VO associated to  node $n_h$.

\paragraph{Users.}

The process model for users, denoted $\userprocess(\til{c},\encp{\mathtt{S}}{}^{t,e}, y)$, is parametric on a tuple of user credentials $\til{c}$, a task process $\encp{\mathtt{S}}{}^{t,e}$ (explained above), and a name $y$, which is used to access a grid node (an instance of process $\approcess(y, \til{d})$). Process $\userprocess(\til{c}, \encp{\mathtt{S}}{}^{t,e}, y)$ interacts with node process $\approcess\langle y, \til{d} \rangle$ in order to authenticate to the grid, create a user proxy, and submit/monitor her task. More precisely, we have:
\begin{eqnarray*}
 \userprocess(\til{c}, \encp{\mathtt{S}}{}^{t,e}, y)  &\myeqdef&  (\nu\ u) (\overline{y}\left\langle \til{c},u \right\rangle.u(ch_1, -, m).\\
& &  \qquad \quad \texttt{if}\ [m=\mathtt{ok}]  \ \texttt{then}\ \overline{ch_1}.ch_1(a).\overline{ch_1}\left\langle \encp{\mathtt{S}}{}^{t,e} \right\rangle.ch_1(g).\monitorprocess \langle \til{c},g,a,y,\textsf{P}_\mathtt{S} \rangle  \\
                                         & &  \qquad\quad \texttt{else}\ \zero) 
\end{eqnarray*}

\noindent Above, the first output on $y$ represents an authentication request
against a  service deployed 
at $\approcess(y, \til{d})$. 
This service 
returns name $\mathtt{ok}$ (resp. $\mathtt{denied}$) if the authentication is successful (resp. failed). 
We write $u(ch_1, -, m)$ to denote a reception of three arguments along $u$, in which the second one is not relevant.
Name $ch_1$ is a private name communicated by  $\approcess(y, \til{d})$: this enables the interference-free communication between user process and a subprocess of the grid node process.
Also, $ch_1$ is used for user proxy creation and task submission: proxy creation is requested by an output signal on $ch_1$; 
then, a name $a$ (to be used to access the user proxy) is received on $ch_1$; 
subsequently, the task can be sent: this is represented by the (higher-order) output prefix $\overline{ch_1}\langle \encp{\mathtt{S}}{}^{t,e} \rangle$. 
Once the task has been sent,
a channel associated to the log of the submitted task is received on $ch_1$, 
and 
 process $\monitorprocess(\til{c},g,a,y,\textsf{P})$ is launched: it 
abstracts the user interaction with her access point for monitoring the task just submitted. 
The last parameter for $\monitorprocess$, process $\textsf{P}_\mathtt{S}$,
specifies the user behavior that is executed upon reception of the final result of her task.
 Such a process may correspond to, e.g., a  query that stores  such a result into a remote database.

\paragraph{Middleware.}
The middleware is represented as the  composition of access point processes $\approcess(y, \til{d})$. For each VO in the grid, there are some instances of access point processes associated to it.
An instance of $\approcess(y, \til{d})$ interacts with an instance of $\userprocess(\til{c}, \encp{\mathtt{S}}{}^{t,e}, y)$ for authentication purposes, user proxy creation, and task submission/monitoring, as just explained. Then, process $\approcess(y, \til{d})$ launches a process $\prxhdlprocess(ce,\til{d})$, given in Fig~\ref{fig:middel}, which interacts with the user proxy process.

\begin{figure}[t!]
\begin{align*}
\prxhdlprocess(ce, \til{d})\myeqdef & ~~ce(\til{k}, m,a,g).(\prxhdlprocess\langle ce, \til{d}\rangle\ \para\ \\
                               &  \qquad \qquad (\nu c,b,f)(\prod\limits_{d_i \in\  \til{d}}\searchprocess^{\til{k}}\langle c,f,d_i \rangle\ \para \ \accprocess\langle c,f,b \rangle\ \para \\
                               &  \qquad \qquad \qquad \qquad b(d_1,\ldots, d_{\sigma}).\sum\limits_{j\in 1 \ldots\sigma}\overline{d_j}\left\langle \til{k}, m, a,g \right\rangle))
\end{align*}
\caption{Process $\prxhdlprocess(ce, \til{d})$, part of the middleware, interacts with the user proxy process.\label{fig:middel}}
\end{figure}
Process $\prxhdlprocess(ce, \til{d})$ is parametric on (i) name $ce$, which is used to receive the task requirements from 
the  user proxy process; and (ii) tuple $\til{d}$, which contains the names associated to the ADs of the VO of the access point.
Once $\prxhdlprocess(ce, \til{d})$ has received on $ce$ the tuple $\til{k}$ which represents the descriptors of the required grid resources, it selects the appropriate ADs for the requested resources. We abstract this selection by processes $\searchprocess^{K}$ and $\accprocess$.
 Given a tuple/set  of resources descriptors $K$, each instance of process $\searchprocess^{K}$ searches among the resources shared by an AD with resources satisfying the requirements in $K$. Once a suitable AD has been found, $\searchprocess^{K}$ sends the access channel of that AD to $\accprocess$, which records all such access channels. Once all instances of $\searchprocess^{K}$ have completed the search, $\accprocess$ sends such ADs to process $\prxhdlprocess(ce, \til{d})$ along name $b$. Then, $\prxhdlprocess(ce, \til{d})$ non-deterministically selects an AD.

\paragraph{Administrative domains.}
As mentioned above, an AD is represented as process $\adprocess(d)$, which consists of the parallel composition of processes in charge of receiving, queuing, and attending task execution requests. Also, $\adprocess(d)$ comprises process models of resources and resource proxies (see below).
For the sake of space, we only present the process $\assignprocess(b,d,ch)$, which is in charge of assigning the appropriate resources for the subtasks assigned to the AD.
This process, given in 
Fig.~\ref{fig:assign}, 
 is parametric on channels $b,d$, and $ch$: it extracts a request of the queue through channels $b$ and $c$, and proceeds to attend it. Then, $\assignprocess(b,d,ch)$ interacts with the local resource manager process through channels $ch$, $ans_1$, and $ans_2$ in order to determine the resources for the request. If appropriate resources for the request are available then $\assignprocess(b,d,ch)$ receives in $ans_1$ the access channels of the resource proxies and forwards them to the user proxy through name $p$. 
Otherwise, if there are no resources then $\assignprocess(b,d,ch)$ receives an input in $ans_2$ and sends  the request back to the queue.


\begin{figure}[t!]
\begin{tabular}{lcl}
$\assignprocess(b,d,ch)$  &$\myeqdef$& $(\nu\ n,c)(\overline{b}\left\langle n,c \right\rangle.(c(k_1,\ldots,k_{\zeta},m,p,g,b'). $\\
                          & & $\qquad \qquad (\nu\ o, ans_1, ans_2)$\\
                          & & $~~ \qquad \qquad (\overline{ch}\left\langle k_1,\ldots,k_{\zeta}, ans_1, ans_2 \right\rangle. $\\
                          & & $~~\qquad \qquad \qquad (ans_1(cr_1,\ldots, cr_{\zeta}).$\\
                          & & $~~\qquad \qquad \qquad  \qquad \overline{p}\left\langle k_1, cr_1,..., k_{\zeta},cr_{\zeta}, m, o \right\rangle.\assignprocess\langle b',d,ch \rangle $\\
                          & & $~~\qquad \qquad \qquad\ + $\\
                          & & $~~\qquad \qquad \qquad ans_2.\overline{d}\left\langle k_1,\ldots,k_{\zeta},m,p,g \right\rangle.\assignprocess\langle b',d,ch \rangle)$\\
                          & & $~~\qquad \qquad  \para o(X).X)$\\ 
                          & & $\qquad \qquad  +\ n.\assignprocess\langle b,d,ch \rangle))$
\end{tabular}
\caption{Process $\assignprocess(b,d,ch)$ assigns appropriate resources for the subtasks assigned to the AD.\label{fig:assign}}
\end{figure}


Observe how also
$\assignprocess(b,d,ch)$ features higher-order communication in its interaction with 
task process $\encp{\mathtt{S}}{}^{t,e}$. In fact, using a higher-order process communication on name $o$ (not shown), 
the task process 
$\encp{\mathtt{S}}{}^{t,e}$ is expected to send a job to $\assignprocess(b,d,ch)$---which is denoted by process variable $X$. As soon as the reception on $o$ takes place, process $\assignprocess(b,d,ch)$ will execute the involved job.

\paragraph{User and Resource Proxies.}
We represent user proxies as instances of a process which receives the requirements of the subtasks of the user task process $\encp{\mathtt{S}}{}^{t,e}$ and submits such requirements to an access point process. 
Moreover, a user proxy process interacts with process $\assignprocess(b,d,ch)$ which sends it the channels of the resource proxies of assigned resources. Finally, the user proxy process communicates with resources proxies process in order to authenticate and obtain the direct access to resources.
Resource proxies are abstracted as a process which interacts with its associated resource process and instances of user proxy process. The interaction with its associated resource process allows the resource proxy to keep track of the state of the resource, as a resource notifies its proxy when a task has been completed. 


\paragraph{Other components.} In addition to the components described above, 
our process models also includes representations for 
other components in the GC system, namely logs processes, resource processes, and queue processes.
There is a log process for each user task: it is in charge of registering the current state and the result of a task. Middleware processes (access points) interact to read the log when the user process requests it. In fact, processes $\prxhdlprocess(ce, \til{d})$ and $\uprxprocess(ce,p,t,g)$ interact with the log process to register a new state and/or the final result.
Resource processes abstract the behavior of actual grid resources. They interact with resource proxy process and task process $\encp{\mathtt{S}}{}^{t,e}$. 
Finally, the queue process is a process representation of a queue structure. There is a queue process for each AD, which is used to store the subtasks requests of resources assigned to the AD.

\section{Formalizing a Representative Grid Scenario}
\label{sec:example}
We now illustrate our formal model by instantiating it with 
the scenario presented in Sect.~\ref{sec:grid} (see also Fig.~\ref{fig:samplegrid}).
The following table summarizes some of the corresponding base sets:
\begin{center}
{\small
\begin{tabular}{rl|l}
& Base Set & Description \\ \hline
$D$&$=\{d_1,d_2, d_3\}$ & $\ $Administrative domains\\
$U$&$=\{u_1, u_2\}$ & $\ $Users\\
$V$&$=\{v_1, v_2\}$ & $\ $Virtual organizations\\
$N$&$=\{n_1, n_2\}$ & $\ $Grid nodes\\
$R$&$=\{r_1,r_2,r_3,r_4,r_5,r_6,r_7, r_8\}$ & $\ $Resources\\
$K$&$=\{k_1,k_2, k_3\}$ & $\ $Resource descriptors\\
$T$&$=\{\mathtt{T_1},\mathtt{T_2}\}$ &  $\ $Task definitions\\
$S$&$=\{\mathtt{S_1}, \mathtt{S_2}\}$ & $\ $User tasks\\
\end{tabular}
}
\end{center}

For the sake of space, we do not present the static component of the model. 
Still, the description of the scenario given in Sect.~\ref{sec:grid} should provide an intuitive idea of the key valid
relationships between the main grid components. We only highlight the fact that 
user tasks $\mathtt{S_1}$ and $\mathtt{S_2}$ are instances of task definitions $\mathtt{T_1}$ and $\mathtt{T_2}$, respectively.
As for the dynamic
component of the model, following the notation given in Fig.~\ref{table:stat_dyna}, our scenario is represented by the following \hopi process:
\begin{eqnarray}
\gridprocess      &=& (\nu\ y_1, y_2) \big(\encp{u_1, \mathtt{S_1}}{}^{\til{c_1},y_1}\ \para\ \encp{u_2,\mathtt{S_2}}{}^{\til{c_2},y_2}\ \para \nonumber\\
                  &&\ \ \ \ \ \ \ \ \ \ \ \ \ \ \ \ (\nu\ d_1,d_2, d_3) (\encp{n_1}{}^{v_1, y_1, d_1,d_2} \ \para\ \encp{n_2}{}^{v_2, y_2, d_2,d_3} \ \para\ \encp{d_1}\ \para\ \encp{d_2}{}\ \para\ \encp{d_3}{})\big)\nonumber
\end{eqnarray}


where $\omega=2$, $\mu=2$, $\delta=2$,	 and $\eta=2$. By expanding the 
definitions of $\encp{u,\mathtt{S}}{}^{\til{c},y}$, $\encp{n}{}^{v, y, \til{d}}$, and $\encp{d_1}{}$, the above process can be equivalently stated as follows:
\begin{eqnarray}
\gridprocess      &=& (\nu\ y_1, y_2) \big((\nu\ t_1, e_1)\ \userprocess\langle \til{c_1},\encp{\mathtt{S_1}}{}^{t_1,e_1},y_1\rangle\ \para\ (\nu\ t_2, e_2)\ \userprocess\langle \til{c_2},\encp{\mathtt{S_2}}{}^{t_2,e_2},y_2\rangle\ \para \nonumber\\
                  &&\ \ \ \ \ \ \ \ \ \ \ \ \ \ \ \ (\nu\ d_1,d_2, d_3) (\approcess^{v_1}\langle y_1, d_1,d_2 \rangle\ \para\ \approcess^{v_2}\langle y_2, d_2,d_3\rangle  \para \adprocess \langle d_1 \rangle \para \adprocess\langle d_2 \rangle \para \adprocess\langle d_3 \rangle)\big)\nonumber
\end{eqnarray}

To illustrate process evolution, we now describe a particular reduction sequence that originates from $\gridprocess$.
Precisely, we show the interactions that occur when the user $u_1$ accesses the grid for executing task $\mathtt{S_1}$.
Clearly, (concurrent) interactions related to user $u_2$ are also possible, but below
we restrict to comment on the reductions related to the process representation of $u_1$. 

First, we have a sequence of reductions $\gridprocess \wred \textsf{GRID}^{1}$, 
that represents the steps in which
$ \userprocess\langle \til{c_1},\encp{\mathtt{S_1}}{}^{t_1,e_1},y_1\rangle$ interacts with process $\approcess^{v_1}\langle y_1, d_1,d_2 \rangle$
to perform steps of user authentication, proxy creation, and submission of task $\mathtt{S_1}$, as stipulated in the protocol. 
Process $\textsf{GRID}^{1}$ is as follows:
\begin{eqnarray*}
\textsf{GRID}^{1}      &\equiv& (\nu\ y_1, y_2) \big(\monitorprocess \langle \til{c_1},g_{r1},a_1,y_1,\textsf{P} \rangle\ \para\ (\nu\ t_2, e_2)\userprocess\langle \til{c_2},\encp{\mathtt{S_2}}{}^{t_2,e_2},y_2\rangle\ \para \nonumber\\
                  &&\qquad \ \ \ \ \ \ \ \ \ \ \ (\nu\ d_1,d_2, d_3) (\approcess^{1}(\mathtt{S_1})\ \para\ \mathsf{RestSystem^1})\big)
\end{eqnarray*}
where residual processes $\approcess^{1}(\mathtt{S_1})$ and $\mathsf{RestSystem^1}$ are as follows:
\begin{eqnarray*}
\approcess^{1}(\mathtt{S_1})      &\equiv& \encp{\mathtt{S_1}}{}^{t_1,e_1}\ \para\ (\nu\ g_{w1}, ce_1)(\logprocess\langle g_{w1}, g_{r1}, \mathtt{submitted}, \mathtt{null} \rangle\ \para\\
                      & & \ \ \ \ \ \ e_1(\til{r}).\overline{g_{w1}}\langle \mathtt{state}, \mathtt{finished} \rangle.\overline{g_{w1}}\langle \mathtt{result}, \til{r} \rangle\ \para\ \uprxprocess\langle ce_1, a_1, t_1, g_{w1} \rangle\ \para\ \nonumber\\
                      & & \ \ \ \ \ \ \prxhdlprocess\langle ce_1,d_1,d_2 \rangle)\nonumber\\
\mathsf{RestSystem^1}    &\equiv& \approcess^{v_1}\langle y_1, d_1,d_2 \rangle\ \para\ \approcess^{v_2}\langle y_2, d_2,d_3\rangle \ \para\ \adprocess \langle d_1 \rangle\ \para\ \adprocess\langle d_2 \rangle\ \para\ \adprocess\langle d_3 \rangle \nonumber
\end{eqnarray*}

In process $\approcess^{1}(\mathtt{S_1})$ above, 
private name $g_{w1}$ is used by   processes $\uprxprocess\langle ce_1, a_1, t_1, g_{w1} \rangle$ and $\prxhdlprocess\langle ce_1,d_1,d_2 \rangle$ to register the changes in the state of task $\mathtt{S_1}$. 
Name $ce_1$ stands for the private channel on which 
these two processes
may interact.
At this point, we have the  reduction sequence $\textsf{GRID}^{1} \wred \textsf{GRID}^{2}$, which represents  reductions corresponding to the AD selection in the VO and the task submission to such an AD. 
In this case, we assume the AD $d_1$ is selected for the execution of the task. 
Process $\textsf{GRID}^{2}$ is as follows:
\begin{eqnarray*}
\textsf{GRID}^{2}      &\equiv& (\nu\ y_1, y_2) \big(\monitorprocess \langle \til{c_1},g_{r1},a_1,y_1,\textsf{P} \rangle\ \para\ (\nu\ t_2, e_2)\userprocess\langle \til{c_2},\encp{\mathtt{S_2}}{}^{t_2,e_2},y_2\rangle\ \para \nonumber\\
                  &&\quad \ \ \ \ \ \ \ \ \ \ \ \ (\nu\ d_1,d_2, d_3) (\approcess^{2}(\mathtt{S_1})\ \para\ \adprocess^{1} \langle d_1 \rangle\ \para\ \mathsf{RestSystem^2})\big)
\end{eqnarray*}
where   $\approcess^{2}(\mathtt{S_1})$ 
and $\adprocess^{1} \langle d_1 \rangle$
stand for residual processes for the access point and for the representation of AD $d_1$, respectively.
As above, $\mathsf{RestSystem^2}$ stands for the composition of processes for the remaining components. 
In process $\adprocess^{1} \langle d_1 \rangle$, the interaction between the task process and the user proxy process has evolved to $\encp{\mathtt{S^1_1}}{}$ 
 and $\uprxprocess_1$, respectively. Processes $\approcess^{2}(\mathtt{S_1})$ and $\mathsf{RestSystem^2}$ are as follows:
\begin{eqnarray*}
\approcess^{2}(\mathtt{S_1})           &\equiv& \encp{\mathtt{S^1_1}}{}\ \para\ (\nu\ g_{w1}, ce_1)(\logprocess\langle g_{w1}, g_{r1}, \mathtt{queued}, \mathtt{null} \rangle\ \para \nonumber\\
                      & & \ \ \ \ \ \ e_1(\til{r}).\overline{g_{w1}}\langle \mathtt{state}, \mathtt{finished} \rangle.\overline{g_{w1}}\langle \mathtt{result}, \til{r} \rangle\ \para \uprxprocess_1 \ \para\ \\
                      & & \ \ \ \ \ \ \prxhdlprocess\langle ce_1,d_1,d_2 \rangle)\nonumber\\
\mathsf{ResSystem^2}    &\equiv& \approcess^{v_1}\langle y_1, d_1,d_2 \rangle\ \para\ \approcess^{v_2}\langle y_2, d_2,d_3\rangle \ \para\ \adprocess\langle d_2 \rangle\ \para\ \adprocess\langle d_3 \rangle \nonumber
\end{eqnarray*}

At this point, we may infer a reduction sequence 
which abstracts steps of 
resource selection and task execution.
We indeed have $\textsf{GRID}^{2} \wred \textsf{GRID}^{3}$, where process $\textsf{GRID}^{3}$ is as follows:
\begin{eqnarray*}
\textsf{GRID}^{3}      &\equiv& (\nu\ y_1, y_2) \big(\monitorprocess \langle \til{c_1},g_{w1},a_1,y_1,\textsf{P} \rangle\ \para\ (\nu\ t_2, e_2)\userprocess\langle \til{c_2},\encp{\mathtt{S_2}}{}^{t_2,e_2},y_2\rangle\ \para \nonumber\\
                  &&\ \ \ \ \ \ \ \ \ \ \ \ (\nu\ d_1,d_2, d_3) (\approcess^{3}(\mathtt{S_1})\ \para\ \adprocess^2(\mathtt{S_1})\ \para\ \mathsf{ResSystem^2})\big)
\end{eqnarray*}

where $\approcess^{3}(\mathtt{S_1})$ corresponds to residual process for the access point; process $\adprocess^2(\mathtt{S_1})$
is its analogous for the representation of AD $d_1$. 
While process $\encp{\mathtt{S^2_1}}{} \equiv \overline{e_1}\langle \til{res_1} \rangle$, 
process $\approcess^{3}(\mathtt{S_1})$  
is as follows:
\begin{eqnarray*}
\approcess^{3}(\mathtt{S_1})           &\equiv&  \encp{\mathtt{S^2_1}}{}\ \para\ (\nu\ g_{w1}, ce_1)(\logprocess\langle g_{w1}, g_{r1}, \mathtt{running}, \mathtt{null} \rangle\ \para\\
                      & & \ \ \ \ \ \ e_1(\til{r}).\overline{g_{w1}}\langle \mathtt{state}, \mathtt{finished} \rangle.\overline{g_{w1}}\langle \mathtt{result}, \til{r} \rangle\ \para \uprxprocess\langle ce_1, a_1, t_1, g_{w1} \rangle \rangle\ \para\ \nonumber\\
                      & & \ \ \ \ \ \ \prxhdlprocess\langle ce_1,d_1,d_2 \rangle)\nonumber
\end{eqnarray*}

Process $\encp{\mathtt{S^2_1}}{}$ stands for the residual process for the task process of user $u_1$; it notifies its completion through channel $e_1$. We obtain the reduction sequence $\textsf{GRID}^{3}\wred \textsf{GRID}^{4}$ after some reductions corresponding to task completion and log registering. Process $\textsf{GRID}^{4}$ is as follows:
\begin{eqnarray*}
\textsf{GRID}^{4}      &\equiv& (\nu\ y_1, y_2) \big(\monitorprocess \langle \til{c_1},g_{w1},a_1,y_1,\textsf{P} \rangle\ \para\ (\nu\ t_2, e_2)\userprocess\langle \til{c_2},\encp{\mathtt{S_2}}{}^{t_2,e_2},y_2\rangle\ \para \nonumber\\
                  &&\ \ \ \ \ \ \ \ \ \ \ \ (\nu\ d_1,d_2, d_3) (\approcess^{4}(\mathtt{S_1})\ \para\ \adprocess^3(\mathtt{S_1}\ \para\ \mathsf{RestSystem^2})\big)
\end{eqnarray*}
where $\approcess^{4}(\mathtt{S_1})$ is as follows:
\begin{eqnarray*}
\approcess^{4}(\mathtt{S_1})           &\equiv& (\nu\ g_{w1}, ce_1)(\logprocess\langle g_{w1}, g_{r1}, \mathtt{finished}, \til{res_1} \rangle\ \para\ \nonumber\\
                      & & \ \uprxprocess\langle ce_1, a_1, t_1, g_{w1} \rangle \rangle\ \para\ \prxhdlprocess\langle ce_1,d_1,d_2 \rangle)
\end{eqnarray*}
At last, 
we may infer the reduction sequence
$\textsf{GRID}^{4}\wred \textsf{GRID}^5$, 
where  
 process $\textsf{GRID}^5$ defined as 
\begin{eqnarray*}
\textsf{GRID}^5      &\equiv& (\nu\ y_1, y_2) \big(\textsf{P}\langle \til{res_1} \rangle\ \para\ (\nu\ t_2, e_2)\userprocess\langle \til{c_2},\encp{\mathtt{S_2}}{}^{t_2,e_2},y_2\rangle\ \para \\
                  &&\ \ \ \ \ \ \ \ \ \ \ \ \ \ \ \ (\nu\ d_1,d_2, d_3) (\approcess^{4}(\mathtt{S_1})\ \para\ \adprocess^3(\mathtt{S_1})\ \para\ \mathsf{RestSystem^2})\big)\nonumber
\end{eqnarray*}
and 
where process
$\textsf{P}\langle \til{res_1} \rangle$ denotes an unspecified, parameterized process that is to be executed by the user monitor with the task result $\til{res_1}$. 

\section{Future Work}
\label{sec:future}

The process model of GC systems presented here 
describes basic interactions among grid main components, abstracting and enforcing essential 
static and dynamic properties of such systems. 
Establishing an operational  
correspondence result connecting the invariants in the static description and the \hopi reductions of the dynamic representation is part of ongoing work.
We conjecture that \hopi processes representing the dynamic part preserve by construction the invariants  defined by the static part. 
Slightly more formally, we conjecture that if process $P$ respects the static invariants, and $P \pired PÕ$ then either (a)~$P'$ preserves the static invariants, or (b)~there is a $P''$ such that $P' \wred P''$ and $P''$ preserves the static invariants. 
One of the challenges in the proof consists in giving a unified treatment to all invariants. 

Our current model does not take into account certain aspects typical of GC infrastructures, such as time.
Still, as already mentioned, we think our current model is already a good basis for extensions: the inherent compositionality of process specifications should ease 
orthogonal improvements and refinements. 
In this sense, as future work, we plan to refine the model with locations (i.e., computation sites) and process failures. 
To this end, an initial approach would be using a calculus of \emph{adaptable processes}~\cite{DBLP:journals/corr/abs-1210-6379}, which enables to incorporate forms of runtime   adaptation over located, interacting processes. 


A strong motivation for pursuing a process calculi model of GC systems is that of exploiting the proof techniques over processes (behavioral equivalences, type systems) so as to reason about grid systems.
That is, we would like to explore how our process model allows us to 
reason about correctness properties of GC systems. This involves, for instance, 
exploiting our model's compositionality and well-established theories of behavioral equivalence to reason about arbitrary behaviors in the grid setting.
Also, we would like to reason about 
task termination and resource delivery in the grid setting. These properties are intrinsically related to reachability problems, and to 
issues of deadlock- and cycle-detection. We believe that a process calculi model offers a suitable basis also for investigating such problems.


\paragraph{Acknowledgments}
We are grateful to Carlos Olarte and to the DCM'13 reviewers for their detailed comments and remarks.
Also, exchanges  with John Sanabria 
were crucial to improve 
our understanding of grid computing platforms.
The research of first author has been supported by COLCIENCIAS 
(the Colombian Agency for 
Science, Technology and Innovation) under a 
``Francisco Jos\'e de Caldas'' doctoral scholarship (Convocatoria 494/2009).
The research of the second author was supported by the Portuguese Foundation for Science and Technology (FCT) under grants SFRH / BPD / 84067 / 2012 and CITI.


\bibliographystyle{eptcs}
\bibliography{bibliografia}

\end{document}